\documentclass[aps,pre,reprint,superscriptaddress]{revtex4-2}

\usepackage[dvips]{graphicx}
\usepackage{bm}
\usepackage{color}

\begin{document}

\preprint{accepted to PRE}

\title{Laser astrophysics experiment on the amplification of magnetic
  fields by shock-induced interfacial instabilities} 


\author{Takayoshi Sano}
\email{sano@ile.osaka-u.ac.jp}
\affiliation{Institute of Laser Engineering, Osaka University, Suita,
  Osaka 565-0871, Japan} 

\author{Shohei Tamatani}
\affiliation{Institute of Laser Engineering, Osaka University, Suita,
  Osaka 565-0871, Japan} 

\author{Kazuki Matsuo}
\affiliation{Institute of Laser Engineering, Osaka University, Suita,
  Osaka 565-0871, Japan} 

\author{King Fai Farley Law}
\affiliation{Institute of Laser Engineering, Osaka University, Suita,
  Osaka 565-0871, Japan} 
\affiliation{Department of Earth and Planetary Science, The University
  of Tokyo, Hongo, Bunkyo-ku, Tokyo 113-0033, Japan} 

\author{Taichi Morita}
\affiliation{Faculty of Engineering Sciences, Kyushu University,
  Kasuga, Fukuoka 816-8580, Japan}  

\author{Shunsuke Egashira}
\affiliation{Institute of Laser Engineering, Osaka University, Suita,
  Osaka 565-0871, Japan} 

\author{Masato Ota}
\affiliation{Institute of Laser Engineering, Osaka University, Suita,
  Osaka 565-0871, Japan} 

\author{Rajesh Kumar}
\affiliation{Institute of Laser Engineering, Osaka University, Suita,
  Osaka 565-0871, Japan} 

\author{Hiroshi Shimogawara}
\affiliation{Institute of Laser Engineering, Osaka University, Suita,
  Osaka 565-0871, Japan} 

\author{Yukiko Hara}
\affiliation{Institute of Laser Engineering, Osaka University, Suita,
  Osaka 565-0871, Japan} 

\author{Seungho Lee}
\affiliation{Institute of Laser Engineering, Osaka University, Suita,
  Osaka 565-0871, Japan} 

\author{Shohei Sakata}
\affiliation{Institute of Laser Engineering, Osaka University, Suita,
  Osaka 565-0871, Japan} 
\affiliation{Administration and Technology Center for Science and
  Engineering, Technology Management Division, Waseda University,
  Okubo, Shinjyuku-ku, Tokyo 169-8555, Japan}

\author{Gabriel Rigon}
\affiliation{LULI, CNRS, CEA, {\'E}cole Polytechnique, UPMC,
  Universit{\'e} Paris 06, Sorbonne Universit{\'e}, Institut
  Polytechnique de Paris, F-91128 Palaiseau Cedex, France}   
\affiliation{Department of Physics, Nagoya University, 
  Chikusa-ku, Nagoya, Aichi 464-8602, Japan} 

\author{Thibault Michel}
\affiliation{LULI, CNRS, CEA, {\'E}cole Polytechnique, UPMC,
  Universit{\'e} Paris 06, Sorbonne Universit{\'e}, Institut
  Polytechnique de Paris, F-91128 Palaiseau Cedex, France}   

\author{Paul Mabey}
\affiliation{LULI, CNRS, CEA, {\'E}cole Polytechnique, UPMC,
  Universit{\'e} Paris 06, Sorbonne Universit{\'e}, Institut
  Polytechnique de Paris, F-91128 Palaiseau Cedex, France}   

\author{Bruno Albertazzi}
\affiliation{LULI, CNRS, CEA, {\'E}cole Polytechnique, UPMC,
  Universit{\'e} Paris 06, Sorbonne Universit{\'e}, Institut
  Polytechnique de Paris, F-91128 Palaiseau Cedex, France}   

\author{Michel Koenig}
\affiliation{LULI, CNRS, CEA, {\'E}cole Polytechnique, UPMC,
  Universit{\'e} Paris 06, Sorbonne Universit{\'e}, Institut
  Polytechnique de Paris, F-91128 Palaiseau Cedex, France}   
\affiliation{Graduate School of Engineering, Osaka University, Suita,
  Osaka 565-0871, Japan} 

\author{Alexis Casner}
\affiliation{Universit{\'e} de Bordeaux-CNRS-CEA, CELIA, UMR 5107,
  F-33405 Talence Cedex, France}  

\author{Keisuke Shigemori}
\affiliation{Institute of Laser Engineering, Osaka University, Suita,
  Osaka 565-0871, Japan} 

\author{Shinsuke Fujioka}
\affiliation{Institute of Laser Engineering, Osaka University, Suita,
  Osaka 565-0871, Japan} 

\author{Masakatsu Murakami}
\affiliation{Institute of Laser Engineering, Osaka University, Suita,
  Osaka 565-0871, Japan} 

\author{Youichi Sakawa}
\affiliation{Institute of Laser Engineering, Osaka University, Suita,
  Osaka 565-0871, Japan} 

\date{\today}

\begin{abstract}
Laser experiments are becoming established as a new tool for astronomical research that complements observations and theoretical modeling.  
Localized strong magnetic fields have been observed at a shock front of supernova explosions. 
Experimental confirmation and identification of the physical mechanism for this observation are of great importance in understanding the evolution of the interstellar medium. 
However, it has been challenging to treat the interaction between hydrodynamic instabilities and an ambient magnetic field in the laboratory.
Here, we developed an experimental platform to examine magnetized Richtmyer-Meshkov instability (RMI).
The measured growth velocity was consistent with the linear theory, and the magnetic-field amplification was correlated with RMI growth.
Our experiment validated the turbulent amplification of magnetic fields associated with the shock-induced interfacial instability in astrophysical conditions for the first time. 
Experimental elucidation of fundamental processes in magnetized plasmas is generally essential in various situations such as fusion plasmas and planetary sciences. 
\end{abstract}

\maketitle


\section{Introduction \label{sec1}}

The shock-induced interfacial instability, which is called the Richtmyer-Meshkov instability (RMI) \cite{richtmyer60,meshkov69}, under the presence of a magnetic field plays a crucial role in various plasma phenomena in astrophysics, space sciences, and laboratory experiments \cite{nishihara10,zhou21}.
The interaction of supernova shocks with the inhomogeneous magnetized interstellar medium is subject to the RMI, which contributes to enforcing interstellar turbulence \cite{inoue09}.
The amplitude of the turbulence has a critical meaning to affect the following star formation history \cite{hennebelle19}.
The RMI is one of the most severe problems in the implosion process of laser-driven inertial confinement fusion (ICF) \cite{atzeni04}.
Ideal compression is achieved only when the mixing caused by the RMI and other interfacial instabilities, e.g., the Rayleigh-Taylor instability (RTI), is mitigated.
Recently, the application of an external magnetic field has been intensely considered for the suppression of the instabilities and electron heat conduction \cite{perkins17}.
Therefore, the understanding of the magnetohydrodynamic (MHD) evolution of the RMI is an urgent issue to be solved. 

The RMI with an ambient magnetic field has been investigated theoretically and numerically.
There are two fundamental interactions of the RMI with a magnetic field.
One is the amplification of the field due to the turbulent velocities associated with the RMI \cite{sano12,matsuoka17}.
The amplification factor could be more than two orders of magnitude, which makes the RMI turbulence a promising mechanism to interpret strong magnetic fields observed at supernova shocks \cite{uchiyama07}.
The amplification occurs when the initial seed field is weak enough.
If the field strength becomes larger than a critical value, the RMI is suppressed by such a strong magnetic field \cite{samtaney03,wheatley05,sano13}.
Thus, the stabilization of the RMI is another essential interaction.
The critical field strength is estimated by the Alfv{\'e}n number for the RMI \cite{sano13,sano21}, which is defined as the ratio of the growth velocity of the RMI to the Alfv{\'e}n speed.
The experimental validation of the theoretical prediction on the MHD RMI remains the next challenge.

The experimental study of the RMI in fluid and gas dynamics has a long history of many decades \cite{meshkov69,jacobs96,brouillette99}. 
However, it is essential to include two key elements, especially for astrophysical applications: an external magnetic field and a strong shock of high Mach number. 
For this purpose, laser-plasma experiments provide a unique and most suitable platform to realize and examine the details of MHD plasma instabilities.

It is known that evolutionary similarity holds between laser and astrophysical plasmas \cite{ryutov99,ryutov00}.
Therefore, phenomena throughout the vast Universe can be understood from laboratory experiments on a very tiny scale \cite{remington99}.
For instance, the RTI is one of the standard subjects of this field, and many experiments have already been performed in high-energy laser facilities, including the National Ignition Facility \cite{kuranz18,casner19,rigon19}.
The generation of magnetic fields associated with the RTI was observed in laser-plasma experiments \cite{manuel12,gao12,nilson15}.
There are also several studies of the RMI experiments in the absence of ambient magnetic fields \cite{dimonte93,farley99,glendinning03,aglitskiy06}.  
In this work, we have therefore conducted the first laser experiment to investigate the interactions between the RMI and the magnetic field.

The purpose of this paper is to experimentally verify the amplification phenomenon of a magnetic field by the RMI under the condition of a weak seed magnetic field.   
From the viewpoint of magnetic-field amplification, the experimental results obtained by the Vulcan and OMEGA laser have been reported \cite{meinecke14,tzeferacos18}.
In contrast, the originality of our experiment highlights the successful observation of the amplification process in much closer situations to the interstellar medium. 
For example, in their experiment, a turbulent flow is forcibly generated by passing a shock wave through an obstacle, while fluid instability naturally generates turbulence in this work. 
The ability to observe a series of evolutions from linear growth to nonlinear turbulence is another important advantage of our experiment.

The outline of this paper is as follows.  
In Sec.~\ref{sec2}, the experimental setup to measure the growth of the RMI and the amplification of magnetic fields is described.  
The experimental achievements are shown in Sec.~\ref{sec3}, which include the growth velocity of the RMI, the interface velocity, and the evidence of magnetic-field amplification.
In Sec.~\ref{sec4}, the physical interpretation of our findings and future prospects are discussed.  
Finally, the conclusions are summarized in Sec.~\ref{sec5}.

\section{Experimental Setup and Method \label{sec2}}

The experiments were performed using the GEKKO HIPER laser facility at the Institute of Laser Engineering, Osaka University.  
The laser is a neodymium-doped glass system operating at the wavelength of the third harmonic $\lambda_{L} = 351$ nm.  
The laser energies between $E_{L} = 185$ and  725 J were delivered to drive a shock wave in the target using a nominally square pulse of 2.5 ns in duration. 
The laser focal spot of 600 $\mu$m in diameter was smoothed using kinoform phase plates \cite{skupsky89}.  
Then the effective laser intensities $I_{L}$ are estimated as a few times $10^{13}$ W/cm$^2$ on the target. 
We define the time origin $t = 0$ by the laser timing in the analysis.  

\begin{figure}
\includegraphics[scale=0.85,clip]{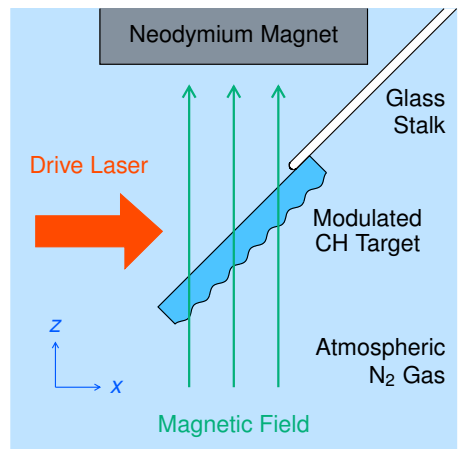}%
\caption{
Side-view sketch of the arrangement for the RMI experiment driven by a laser-induced shock in a weak ambient magnetic field.
The GEKKO laser irradiates a polystyrene foil target in nitrogen gas. 
A permanent magnet applies the initial seed field at the target position.
\label{fig1}}
\end{figure}

The experimental setup was designed to be as simple as possible (see the sketch in Fig.~\ref{fig1}) to obtain the evidence of magnetic-field amplification by the RMI.   
The GEKKO laser irradiated a polystyrene (CH) foil with a thickness of 50 $\mu$m. 
The surface modulation was applied to the rear side of the foil in advance. 
The modulation shape was imprinted by the heat press on a wavy pattern of tungsten mold.
The averaged wavelength of the modulation is $\lambda = 154 \pm 4$ $\mu$m, and the measured amplitude is typically $\psi_0 = 8.8 \pm 0.5$ $\mu$m.
The foil size is $1.4 \times 0.4$ mm, so the width of the target is smaller than the laser spot.  
The foil target alone was held by a glass stalk in the target chamber filled with nitrogen gas (N$_2$).  
For the external magnetic field, we placed a neodymium magnet at 6 mm above the target. 
The size of the cylindrical magnet is 12 mm in diameter and 16 mm in height. 
The magnetic field strength is 0.63 T at the surface of the magnet, and then the seed field $B_{\rm ext}$ is about 0.08 T at the target position. 
{\color{black} 
The magnetic field variation within the target size is at most 6\%, and the nonuniformity has little influence on the later evolution.}
The angle between the target surface and the external magnetic field is about 45 degrees in our setup.

The boundary between the rear side of the target ($\rho_{\rm CH} = 1.0$ g/cm$^{3}$) and N$_2$ gas creates a modulated contact discontinuity. 
When the laser-driven shock reaches the rear surface, the interface is subject to the RMI. 
{\color{black}
The gas pressure was $(6.7 \pm 0.3) \times 10^2$ Pa, in which the mass density of the nitrogen is estimated as $\rho_{{\rm N}_2} \approx 8.3 \times 10^{-6}$ g/cm$^{3}$.}
This case is the heavy-to-light configuration with a huge density jump $\rho_{{\rm N}_2} / \rho_{\rm CH} \ll 1$, and the absolute value of the Atwood number, ${\rm At} = ( \rho_{{\rm N}_2} - \rho_{\rm CH} ) / ( \rho_{{\rm N}_2} + \rho_{\rm CH} )$, is almost unity. 

The induction coil probe (also known as the B-dot probe) was used to measure time-varying magnetic fields according to Faraday's law of induction \cite{everson09}. 
Three orthogonal components of the magnetic field were detected with the independent coils.   
An oscilloscope recorded the electromotive force in the voltage induced when the magnetic flux within the coil changes in time.
The oscilloscope had 1 GHz bandwidth with a sampling time interval of 100 ps (10 GHz), whereas the frequency spectra of the magnetic field considered in this work are at 1--30 MHz.
The B-dot probe has a nearly linear response in this frequency range. 
The same technique was adopted in similar experiments at the LULI2000 \cite{gregori12} and Vulcan laser facilities \cite{meinecke14,meinecke15}. 
In order to capture the magnetic field moving with the turbulent interface, the B-dot probe should be set along the direction of the plasma flow blown out from the rear surface. 
The location of the probe in our setup was 4.2 cm away from the laser focal spot in the direction perpendicular to the foil surface. 
We also performed an off-axis measurement with the same probe for comparison, which was 52 degrees offset from the plasma flow axis. 

An extensive array of visible diagnostics has been implemented on the GEKKO laser facility for various experiments \cite{koenig17,kuramitsu18,morita19,michel20}.  
Besides the drive beams, a probe YAG laser at 532 nm is available in the direction perpendicular to the shock propagation.
The energy of the probe laser is a few mJ, and the pulse duration is about 10--15 ns.  
The time evolution of the interface between the CH target and N$_2$ gas was observed by a simple shadowgraph coupled with three cameras with a time-gated intensified CCD (ICCD) detector, which allow multiple snapshots of the silhouette against the background of the probe light.  
The exposure time of the cameras for the shadowgraph was 200, 250, and 1600 ps, which can be regarded as instantaneous compared to the growth timescale of the RMI.  
Optical pyrometry for self-emission from the shocked plasma was taken by another ICCD camera with an exposure time of 5 ns.  
The observation bandwidth for self-emission is 10 nm around the wavelength of 450 nm.  

The streaked diagnostics were also implemented for the shadowgraph and self-emission. 
The streak cameras enable the measurement of trajectories of the interface and shock front in each shot.   
The slits for the streaked images were aligned parallel to the plasma flow direction.

The seed magnetic field must be weak enough not to suppress the RMI growth. 
The criterion for the suppression is given by the Alfv{\'e}n number evaluated as ${\rm Al} = {\delta v} / {v_A} \lesssim 1$ \cite{sano13,sano21}, where $v_A = B_{\rm ext} / (\mu \rho)^{1/2}$ is the Alfv{\'e}n speed and $\mu$ is the permeability. 
The slower value of the Alfv{\'e}n speed at the interface plays a decisive role in the suppression process, and thus the CH density should be considered.
The growth velocity of the RMI in our experiment is anticipated to be about $\delta v \sim 3$ km/s.   
The required field strength $B_{\rm crit}$ for the suppression is then given by
\begin{equation}
B_{\rm crit} \; {\rm [T]} \sim 100 
\left( \frac{\rho}{ 1 \; {\rm g/cm}^3} \right)^{1/2}
\left( \frac{\delta v}{3 \; {\rm km/s}} \right)
 \;,
\label{eq:bcrit}
\end{equation}
where the permeability in the vacuum $\mu = \mu_0$ is assumed for simplicity. 
The stabilization by the magnetic field is more efficient as the growth velocity becomes slow or the target density is low. 
In our case, the field strength at the target was far below the critical value, i.e., $B_{\rm ext} \ll B_{\rm crit}$.
Therefore, the RMI could take place, and the field amplification by the turbulent motion is strongly expected. 

\section{Experimental Results \label{sec3}}

\subsection{Growth of interfacial instabilities}

\begin{figure*}
\includegraphics[scale=0.85,clip]{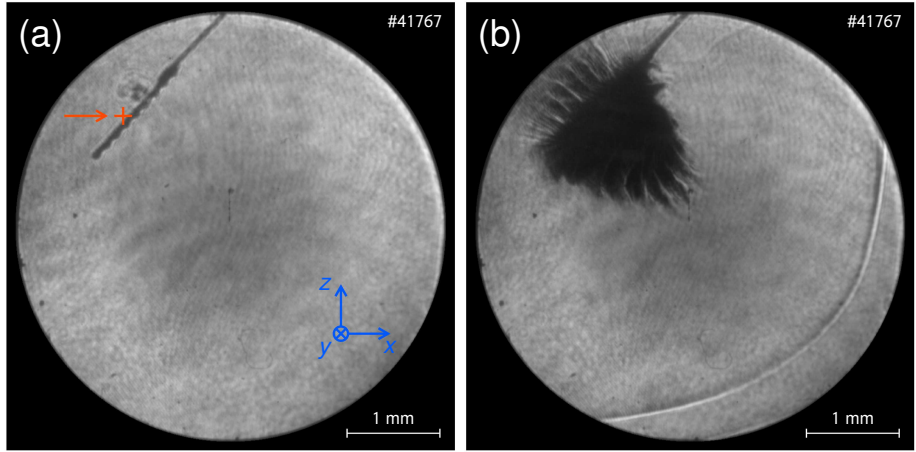}%
\caption{
Optical shadowgraph images of the target in shot No. 41767 (a) before the shot and (b) 40 ns after the shot.
A modulated CH target was used in this shot so that the rear surface is subjected to the RMI.
In the later evolutionary stage shown by (b), a fluctuated contact surface and a smooth transmitted shock emerged as a shadow. 
The field of view is 4.63 mm in diameter.
The red arrow and mark in (a) denote the drive laser injection and the center of the laser focal spot, respectively.
The indicated coordinate is for the three-axis induction coil probe.
\label{fig2}}
\end{figure*}

\begin{figure*}
\includegraphics[scale=0.85,clip]{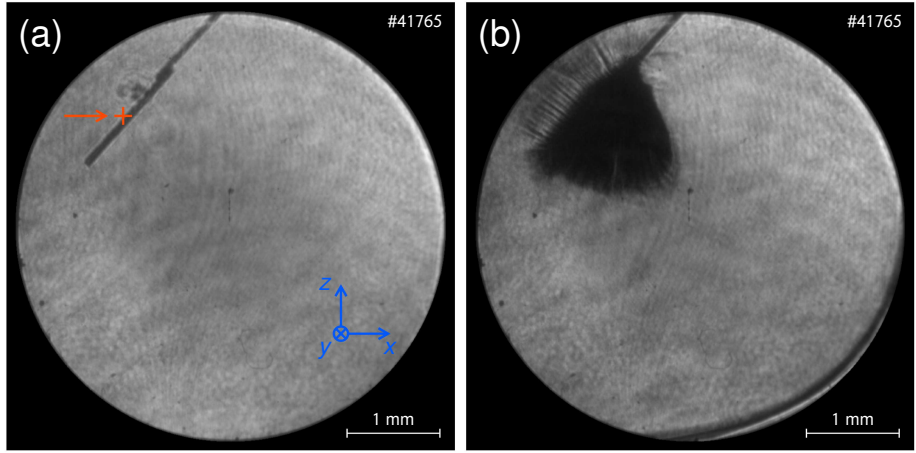}%
\caption{
Optical shadowgraph images of the target in shot No. 41765 (a) before the shot and (b) 40 ns after the shot.
A flat CH target was used in this shot so that the rear surface is stable for the RMI.
The indicated marks are the same as in Fig.~\ref{fig2}.
\label{fig3}}
\end{figure*}

The growth of the RMI was observed through the optical shadowgraph.
Figures~\ref{fig2}(a) and \ref{fig2}(b) show snapshot images of the polystyrene foil target before the shot and 40 ns after the shot, respectively.
The field of view of the instrument is 4.63 mm in diameter.
In the figure, the GEKKO laser comes horizontally from the left to the target, and the red mark indicates the center of the focal spot. 
We define the Cartesian coordinate of this system in which the $x$ axis is parallel to the laser injection, and the $z$ direction is upward on the image. 
The incident angle of the laser is 45 degrees to the target surface in our setup (see Fig.~\ref{fig1}).  
For this shot, the laser energy delivered to the target was $E_{L} = 201$ J.
The intensity corresponds to $I_{L} = 9.0 \times 10^{12}$ W/cm$^2$, where the influence of the incident angle and the conversion efficiency of the phase plate ($\approx 45\%$) are taken into account. 
The silhouette of the CH foil shone by the probe YAG laser traces the contact surface with the ambient N$_2$ gas.
The spatial resolution evaluated at the target edge is $2.7 \pm 0.4$ pixels. The pixel size of this image corresponds to about 5.06 $\mu$m.
Then, in Fig.~\ref{fig2}(a), a sinusoidal pattern on the rear surface with about nine wavelengths long is marginally resolved.

A shock wave produced immediately after the laser irradiation propagates towards the rear side of the target, and finally interacts with the modulated interface. 
The shock velocity in the CH foil is about 20 km/s, which is simulated by the radiation-hydrodynamic code MULTI \cite{ramis88} assuming $I_{L} = 10^{13}$ W/cm$^2$.
Then the shock transit time through the CH target would be within three nanoseconds.

The growth of the RMI enhanced the amplitude of the modulation, which is observed clearly in Fig~\ref{fig2}(b).
This image was taken at $t = 40$ ns, and the exposure time of the camera was 250 ps. 
As shown, the rear side of the target is severely distorted at this time.
The wavelength of the finger-like structure is nearly consistent with the initial wavelength of the modulation $\lambda \approx 150$ $\mu$m. 
Thus, the structure suggests it is an outcome of unstable growth of the initial perturbation. 
The spike-top does not show the mushroom shape in our experiment, which is reasonable because the straight finger-like structure is the characteristic feature of the RMI with a large density jump \cite{matsuoka06}.  
The finger-length from the peak to the valley is about $2 \psi \sim 340$ $\mu$m, which is much larger than the initial amplitude of $\psi_0 \approx 9$ $\mu$m.
Assuming linear growth of the amplitude with time, the growth velocity estimated from Fig.~\ref{fig2} is $\langle \delta v \rangle \approx 4.0$ km/s on average.

The shadowgraph image captured the stable surface of the transmitted shock because it is sensitive to the second derivative of the column density \cite{hutchinson02}.
{\color{black}
Since the corrugation of the transmitted shock front dies away quickly after it propagates the order of the fluctuation wavelength \cite{wouchuk01b,coboscampos17}, the observed shock surface is smooth by contrast to the interface.}
The shock speed is faster than the interface velocity, so that the shock front locates far beyond the contact discontinuity.
The distance to the shock front from the original target position is 3.5 mm, which gives an estimation of the average shock velocity of about 86 km/s in the gas.  
{\color{black}
Assuming the nitrogen gas temperature is around 10 eV, or the sound speed is $c_s \approx 8.2$ km/s, the Mach number of the transmitted shock is about 10.
The plasma beta is very large in this situation, so that the Alfven Mach number would be much larger than 10.}

The phase reversal of the initial modulation is a characteristic behavior of the RMI when the rarefaction is reflected \cite{shigemori00}.
This feature is identified by the shadowgraph image at the earlier phase of $t = 25$ ns in a similar shot. 
When the shock hits the rear surface, the reflected rarefaction travels back to the front surface. 
The trace left by the modulated rarefaction wave is seen in the shadow of the ablation plasma that exhibits an orderly periodic pattern with the initial modulation wavelength.

The evidence of the instability is also confirmed by comparing with the result of a flat foil target, which is displayed in Fig.~\ref{fig3}. 
The rear surface of the target has no initial modulation [see Fig.~\ref{fig3}(a)] so that the RMI growth cannot be expected. 
The laser intensity of this shot was $9.2 \times 10^{12}$ W/cm$^2$, which means that the experimental conditions are almost the same as in the shot shown in Fig.~\ref{fig2} except for the foil shape.  
In contrast to the modulated-target case (Fig.~\ref{fig2}), the contact surface is smooth and stable even at $t = 40$ ns.
The wavefront of the contact surface reaches 1.1 mm from the laser spot, which is equivalent to an average interface velocity of $\langle v_i \rangle \sim 26$ km/s.
A stable shock surface is also visible near the edge of Fig.~\ref{fig3}(b).
The shock front position is 3.9 mm, and the average shock velocity is 96 km/s for this case.
{\color{black}
The difference in the shock velocity compared with that in Fig.~\ref{fig2} might be due to the fluctuations in the laser intensity and the ambient gas pressure.}
At the laser ablation side, we can see thin striped structures each at an interval comparable to the thickness of the target. 
Although it could be an indication of the ablative RTI \cite{takabe85,betti98}, the interpretation of this peculiar structure is beyond the scope of this paper.

Since the linear growth velocity of the RMI increases with increasing the Mach number of the incident shock \cite{wouchuk97}, it should depend on the laser energy or intensity. 
We can change the laser intensity by increasing the number of laser beams (e.g., 3, 6, or 9).
In our experiment, the growth velocity was evaluated from three snapshot data taken by the ICCD cameras at different timing during the same shot.
The shadowgraph is limited by the pulse duration of the probe laser.
Then, the time interval of the images mainly was 5 ns, and the full range of the measurement period was 10--15 ns. 
The amplitude of the finger-like fluctuation of the wavelength $\lambda$ was evaluated from the rear surface shape in each snapshot.
We define half of the spike-to-bubble distance as the average amplitude of $\psi$ at that time.
The growth velocities, $\delta v = d \psi / d t$, are obtained by linear fitting of the time profile of the amplitude.
The obtained growth velocity for each shot is listed in Table~\ref{tab1} with its corresponding laser condition.  
The identification of the spike top or bubble bottom is not obvious in some images, which is reflected by the relatively larger error in $\delta v$. 
{\color{black}
In addition, the timing to evaluate the growth velocity is different for each shot, which could contribute to the data variability.}

\begin{figure}
\includegraphics[scale=0.85,clip]{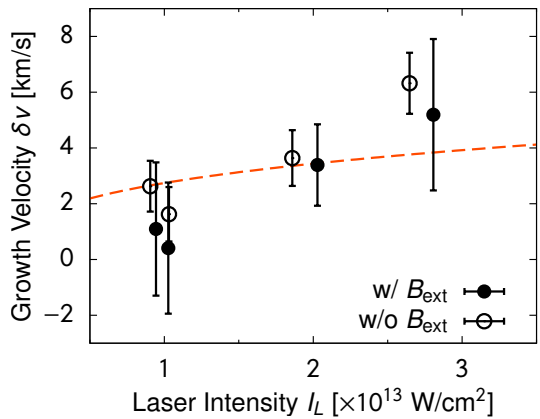}%
\caption{
Growth velocities of the modulation amplitude obtained in the GEKKO-laser experiment.
The closed and open circles are the results of the shots with and without the initial magnetic field applied by a permanent magnet. 
The corresponding data shown in this figure are listed in Table~\ref{tab1}. 
The red dashed curve is the linear growth velocity of the Wouchuk-Nishihara formula \cite{wouchuk97} using an experimentally-obtained interface velocity given by Eq.~(\ref{eq:vint}).
Here, the numerical factor is assumed to be $|\xi| = 0.3$ (see Appendix~\ref{seca}).
\label{fig4}}
\end{figure}

As expected, there is a positive correlation between the growth velocity of $\delta v$ and the laser intensity $I_{L}$.  
Experimentally measured growth velocities are displayed as a function of the laser intensity in Fig.~\ref{fig4}. 
The theoretical growth velocity is also shown in the figure (see Appendix~\ref{seca} for the derivation).
In order to study the impact of the initial magnetic field on the RMI growth, we performed the experiments of the modulated target not only with the magnet but also without the magnet. 
The growth velocities in the experiments with and without the seed field are plotted in Fig.~\ref{fig4} as shown by the closed and open circles, respectively. 
There is no systematic difference in the growth velocity caused by the inclusion of the magnet. 
Although the growth velocity is slightly lower for the shots with the external magnetic field, the difference is within the error.
{\color{black} 
The errors in the shots with the initial magnetic field appear to be larger. 
The plasma beta value, which is the ratio of the thermal pressure to the magnetic pressure, is much larger than unity in this experiment.
This fact suggests that the magnetic field is too weak to affect the dynamics of the RMI \cite{sano13,sano21}.
Thus, the measured growth velocities and the errors would be independent of the external magnetic field. }

\begin{table*}
\caption{
Growth velocities of the RMI obtained in the laser experiment with the modulated CH target.
The first three columns are the shot number, laser energy $E_{L}$, and intensity $I_{L}$ of each shot.
For the conversion from the laser energy to intensity, we considered the incident angle and the transmittance of the phase plate.
The growth velocities $\delta v$ (column 4) are estimated from multiple shadowgraph images during the measurement period (column 5).
The last two columns are the availability of the magnet and the B-dot probe in the shot.
\label{tab1}}
  \begin{tabular}{lcccccc}
    \hline \hline
    Shot No. & Laser Energy &
    Intensity & Growth Velocity &
    Measurement Period & Magnet & B-dot Probe\\ 
    & $E_{L}$ [J] & $I_{L}$ [W/cm$^2$] & $\delta v$ [km/s] &
    [ns] &  & \\ \hline
    41767 & 201 & 0.90 $\times$ 10$^{13}$ & 2.6 $\pm$ 0.9 & 30 -- 40 & -
    & Yes \\
    40335 & 209 & 0.94 $\times$ 10$^{13}$ & 1.1 $\pm$ 2.4 & 45 -- 55 &
    Yes & - \\
    41742 & 227 & 1.0 $\times$ 10$^{13}$ & - & - & - & Yes \\
    41763 & 228 & 1.0 $\times$ 10$^{13}$ & 0.4 $\pm$ 2.3 & 15 -- 25 &
    Yes & Yes \\
    40334 & 229 & 1.0 $\times$ 10$^{13}$ & 1.6 $\pm$ 1.0 & 45 -- 55 &
    - & - \\
    41744 & 239 & 1.1 $\times$ 10$^{13}$ & - & - & Yes & Yes \\
    40344 & 413 & 1.9 $\times$ 10$^{13}$ & 3.6 $\pm$ 1.0 & 45 -- 55 &
    - & - \\
    40333 & 450 & 2.0 $\times$ 10$^{13}$ & 3.4 $\pm$ 1.5 & 30 -- 40 &
    Yes & -\\
    40345 & 588 & 2.6 $\times$ 10$^{13}$ & 6.3 $\pm$ 1.1 & 45 -- 60 &
    - & - \\
    40342 & 623 & 2.8 $\times$ 10$^{13}$ & 5.2 $\pm$ 2.7 & 30 -- 40 &
    Yes & - \\
    \hline \hline
  \end{tabular}
\end{table*}

\subsection{Comparison with the theoretical growth velocity of RMI}

\begin{figure}
  \includegraphics[scale=0.85,clip]{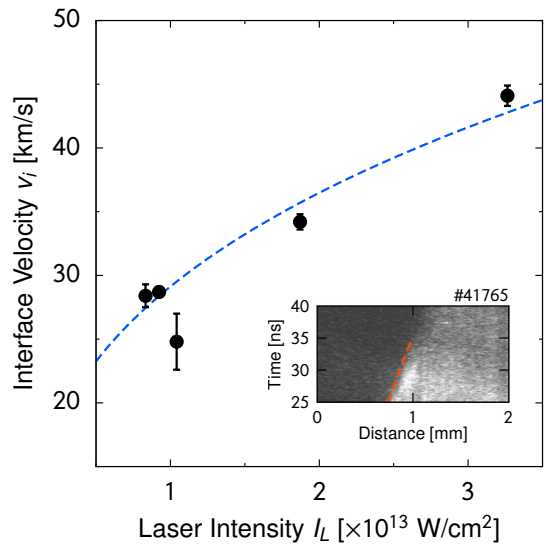}%
\caption{
Interface velocities obtained in the experiment of the flat target with the seed magnetic field.
The corresponding data shown in this figure are listed in Table~\ref{tab2}. 
The blue dashed curve is the power-law fitting of the data given by Eq.~(\ref{eq:vint}). 
Here the interface velocity is fitted by a function $a I_L^b$ where $a$ and $b$ are the fitting parameters.
(inset)
Streaked image of the shadowgraph for shot No. 41765.
The silhouette of the interface between the CH foil and N$_2$ gas is captured in this figure.
The boundary is indicated by the red dashed line.
\label{fig5}}
\end{figure}

Here we will check whether the RMI indeed initiates the enhancement of the modulation amplitude observed in our experiment.
Based on the linear stability analysis of the RMI \cite{richtmyer60,meyer72,wouchuk97}, the growth velocity is correlated to the interface velocity. 
The interface velocity is one of the observable quantities in our experiment.
Then, the theoretical growth velocity inferred from the observed interface velocity must be consistent with the experimentally obtained growth velocity if it is driven by the RMI.

The interface velocity $v_i$ was evaluated in the flat-target shots based on two methods, which namely are from multiple shadowgraph images of different timings and the trajectory of the interface position in streaked images. 
The inset of Fig.~\ref{fig5} is a streaked image of the shadowgraph for a flat-target shot, which traces the interface trajectory through the edge of the shadow.
The interface velocity is extracted by the gradient of the trajectory in the position-time diagram, as shown by the red dashed line in the figure. 
Note that the pulse duration of the probe laser was limited to at most 15 ns so that the streaked shadowgraph is restricted only in this time range. 
The obtained data are listed in Table~\ref{tab2} with the corresponding information of the measured period for each shot.

It is found that the interface velocities exhibit a power-law dependence on the intensity, which is depicted in Fig.~\ref{fig5}.
The fitted function is given by
\begin{equation}
  v_i \; {\rm [km/s]} = \left( 29.1 \pm 0.7 \right)
  \left( \frac{I_{L}}{10^{13} {\rm W/cm}^2}
  \right)^{0.33 \pm 0.04} 
  \;.
\label{eq:vint}
\end{equation}
Here we assume that the decrease of the interface velocity with time is not so significant and then ignore the difference in the measured period.
The ablation pressure has a simple relation with $P_a \propto (I_{L} / \lambda_{L})^{2/3}$ where $\lambda_{L}$ is the laser wavelength \cite{atzeni04}.
The dependence given by Eq.~(\ref{eq:vint}) is consistent with the interpretation that the interface velocity is proportional to the sound speed determined by the ablation pressure $v_i \propto P_a^{1/2}$. 

Based on the linear analysis \cite{wouchuk97}, the growth velocity of the RMI is described as a function of the interface velocity.
Suppose the experimental parameters are given such as the density jump $\rho_{{\rm N}_2} / \rho_{\rm CH} \approx 10^{-5}$, modulation amplitude $\psi_0 / \lambda \approx 0.05$, and isentropic index $\gamma = 5/3$.  
Then, the growth velocity of the Wouchuk-Nishihara (WN) formula \cite{wouchuk97} is expressed as
\begin{equation}
  | v_{\rm wn} |
  \approx  0.094 \left( \frac{| \xi |}{0.3} \right)
  \left( \frac{\psi_0 / \lambda}{0.05} \right) v_i \;,
    \label{eq:vwn}
\end{equation}
{\color{black}
where $\xi$ is a non-dimensional factor obtained from the detailed calculation of the growth velocity (see Appendix~\ref{seca}).
The absolute value of this factor is $| \xi | \lesssim 1$, and of the order of 0.1.
Thus, the analytical growth velocity is given approximately by ten percent of the interface velocity, $| v_{\rm wn} | \sim 0.1 v_i$.}

By substituting the experimental result of $v_i$ [Eq.~(\ref{eq:vint})] into Eq.~(\ref{eq:vwn}), the theoretical prediction of the growth velocity can be estimated.
The obtained $| v_{\rm wn} |$ is drawn by the red dashed curve in Fig.~\ref{fig4}.  
The growth velocities in our experiment are the same order of the theoretical expectation $|v_{\rm wn}|$.
The order-of-magnitude consistency, therefore, implies that the RMI genuinely causes the enhancement of the modulation amplitude in our experiment. 

However, for the higher intensity cases, the experimental data seem to be slightly faster than the theory.
In general, the higher laser intensity produces a laser-driven shock with a higher Mach number. 
In the limit of the high Mach number, the growth velocity becomes much slower than the interface velocity, that is, $| \xi |$ becomes smaller (see Fig.~\ref{fig9} in Appendix~\ref{seca}).
Therefore, the deviation between the experimental $\delta v$ and the theoretical $|v_{\rm wn}|$ is more pronounced in the higher intensity cases.
The contamination of the RTI might be the source of the enhancement of the unstable growth observed in the experiment. 
It is because deceleration of the interface due to the geometrical effect appears in the earlier timescale if the interface velocity is fast or the laser intensity is high. 

\begin{figure}
  \includegraphics[scale=0.85,clip]{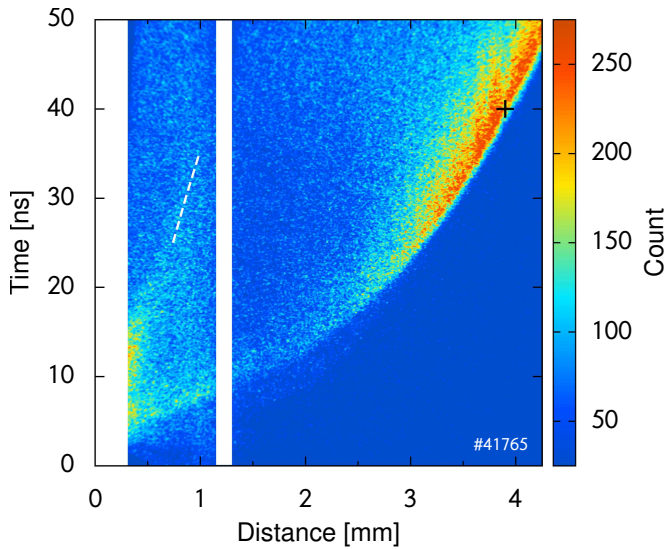}%
\caption{
Streaked image of the self-emission from the shocked gas in shot No. 41765.
The origin stands for the initial target position and the laser timing.
The time profile of the shock velocity is obtained from the trajectory of the shock front. 
The shock-front position obtained from the shadowgraph image [Fig.~\ref{fig3}(b)] is plotted by the black mark at (3.9 mm, 40 ns).
The white dashed line is the interface location measured by the shadowgraph streak image of this shot shown by the inset of Fig.~\ref{fig5}.
The blank data near 1.2 mm is due to the damage of the camera.
\label{fig6}}
\end{figure}

The streaked images of self-emission reveal the long-term evolution of the shock front.
The advantage of the self-emission measurement is to be available in a longer time window up to 50 ns. 
Figure~\ref{fig6} shows a sample image of the streaked self-emission for a flat target case.
The edge of the strong emission traces the shock front, and the gradient in the position-velocity diagram gives the shock velocity.
The shock-front position at $t = 40$ ns is coincident with the shadowgraph image of this shot [Fig.~\ref{fig3}(b)].
Because the shock front travels much further than the laser spot size, the decrease of the shock velocity due to the geometrical effect is not negligible.   
In fact, the shock velocity until $t \sim 10$ ns is about 150 km/s, while it is less than 100 km/s after $t \sim 20$ ns.
The interface velocity could also be decelerated at the later evolutionary stage.
As a reference, the trajectory of the interface is indicated by the white dashed line. 
The effect of the RTI at the decelerating interface is evaluated in Section~\ref{sec4}.

\begin{table*}
\caption{
Interface velocities obtained in the laser experiment with the flat CH target.
The definition of the first three columns is the same as in Table~\ref{tab1}.
The interface velocities $v_i$ (column 4) are estimated from multiple shadowgraph images or the streaked image during the measurement period (column 5) by the fitting of a linear function passing through the origin.
The permanent magnet is set in all the shots listed in this table.
The last column is the availability of the B-dot probe in the shot.
\label{tab2}}
  \begin{tabular}{lcccccc}
    \hline \hline
    Shot No. & Laser Energy &
    Intensity & Interface Velocity &
    Measurement Period & Magnet & B-dot Probe \\ 
    & $E_{L}$ [J] & $I_{L}$ [W/cm$^2$] & $v_i$ [km/s] &
    [ns] &  & \\ \hline
    41747 & 185 & 0.83 $\times$ 10$^{13}$ & 28.4 $\pm$ 0.9 & 30 -- 40 &
    Yes & Yes \\
    41765 & 205 & 0.92 $\times$ 10$^{13}$ & 28.7 $\pm$ 0.4 & 25 -- 35 &
    Yes & Yes \\
    40332 & 232 & 1.0 $\times$ 10$^{13}$ & 24.8 $\pm$ 2.2 & 30 -- 40 &
    Yes & - \\
    40338 & 416 & 1.9 $\times$ 10$^{13}$ & 34.2 $\pm$ 0.6 & 45 -- 55 &
    Yes & - \\
    40341 & 725 & 3.3 $\times$ 10$^{13}$ & 44.1 $\pm$ 0.8 & 30 -- 40 &
    Yes & - \\
    \hline \hline
  \end{tabular}
\end{table*}

\subsection{Amplification of ambient magnetic fields}

The induction coil probe was used to detect the enhancement of a seed magnetic field.
{\color{black}
In order to eliminate the electrostatic component, two electrically independent wires per axis are used, which are twisted together and wound counter-direction \cite{everson09}. 
In this case, an external electric field acts equally on the charges in the coils, and then the voltage of the same polarity arises on each coil-pair. 
However, a magnetic field induces a voltage of opposite polarity.
Therefore, subtracting one from the other gives twice the magnetic field component and cancels out the contribution of the electrostatic component.}

In principle, it is possible to quantify the magnetic field strength using the B-dot probe.
However, the electromagnetic noise is significant in high-intensity laser experiments, and its subtraction is not straightforward. 
Thus, in this analysis, we focus on using the raw signals in voltage of the coil detection and the Fourier spectra.
The Cartesian coordinate of the three-axis probe is depicted in Figs.~\ref{fig2}(a) and \ref{fig3}(a).
In our target configuration, the seed magnetic field would have a dominant component in the $z$ direction.

\begin{figure}
\includegraphics[scale=0.85,clip]{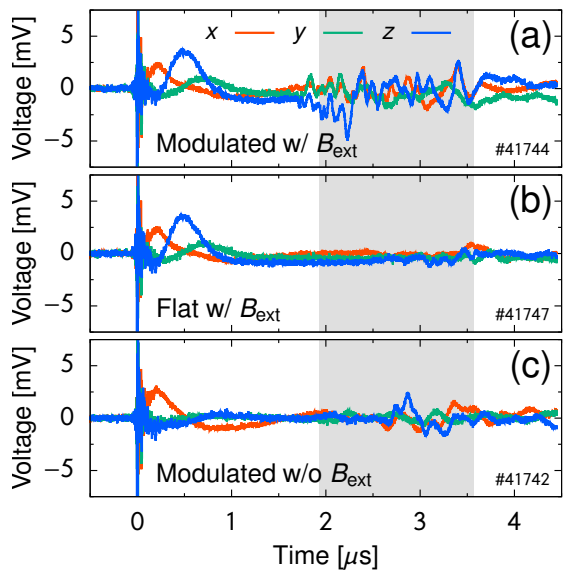}%
\caption{
Time profiles of the signals in voltage detected by the inductive coil probes.
The amplitude indicates the time derivative of the magnetic field strength for (a) a modulated-target case with the magnet, (b) a flat-target case with the magnet, and (c) a modulated-target case without the magnet.
The shot number is indicated at the right-bottom of each panel.
Each component of the signals is shown in different colors.
The later-phase signals in the highlighted period ($1.93 < t$ [$\mu$s] $< 3.57$) are used for the Fourier analysis shown in Fig.~\ref{fig8}.
{\color{black}
The significant amplitude of signal noise at the laser timing $t = 0$ can be seen in all cases.}
  \label{fig7}}
\end{figure}

Two characteristic features have been retrieved from the B-dot data at different times. 
Figure~\ref{fig7} displays the signals from the B-dot probe for three different shots.  
The top and middle panels compare the features based on the target shape.
The bottom one is for the demonstration of the self-generated magnetic field.
The modulated and flat targets are used in Figs.~\ref{fig7}(a) and \ref{fig7}(b), respectively.
Figure~\ref{fig7}(c) is for the case of the modulated target when the initial magnetic field is off.
The large amplitude of the signals appears typically around $t \sim 0.5$ $\mu$s and after $t \sim 2$ $\mu$s.

The transmitted shock seems to contribute to the earlier-phase signals. 
Because the coil probe locates at 4.2 cm away from the laser spot, the arrival time $t \sim 0.5$ $\mu$s implies the plasma velocity is $v \sim 80$ km/s, which is consistent with the observed shock velocity in the gas.
The early-phase B-dot signals in the top two panels are almost identical, but it is different in the case without the magnet. 
Hence, this feature could be caused by the compressional amplification of the seed magnetic field at the shock surface. 

A noticeable difference exists in the later-phase signals in Figs.~\ref{fig7}(a) and \ref{fig7}(b).
After $t \sim 2$ $\mu$s, the modulated-target shot exhibits largely fluctuating signals, while the signals in the flat-target shot are considerably quiet.
The plasma velocity carrying these signals is around 20 km/s or less. 
The laser intensity of these shots was about $I_{L} \sim 10^{13}$ W/cm$^2$ so that the interface velocity is expected to be $v_i \sim 30$ km/s.
Taking into account the velocity decay by the spherical expansion, it is reasonable that the later-phase signals originated from the magnetic field associated with the interface plasma.  
All three components are evenly fluctuating, which means the field direction is randomized in the plasma.
These characteristics of the B-dot signals are explained by the magnetic field in the RMI turbulence, and thus provide clear evidence of the field amplification by the RMI.  

An interesting comparison can be made by using the result of the modulated target without a seed magnetic field, which is shown by Fig.~\ref{fig7}(c).
The later-phase signals in this shot exhibit fewer fluctuations compared with the case of the RMI with a seed magnetic field [Fig.~\ref{fig7}(a)], but slightly more evident than the flat-target case [Fig.~\ref{fig7}(b)].
It is known that the turbulent motions in plasmas could generate magnetic fields via the so-called Biermann battery effect \cite{biermann50}.
The signals around $t \sim 3$ $\mu$s in Fig.~\ref{fig7}(c) indicate such a self-generated field.
Therefore, we can categorize three types of magnetic-field evolutions; (i) a mixture of both amplified and self-generated fields, (ii) no amplification and no self-generation of the field, and (iii) only a self-generated field but no amplification of a seed field.

It should be noticed that no other signal except for the shock signal was detected until 9 $\mu$s, when the B-dot probe was set at the off-axis location of the plasma flow.
The distance from the focal spot to the probe was 7 cm for this case.
The negative detection additionally supports our interpretation that the magnetic-field signals around 2--3.5 $\mu$s in Fig.~\ref{fig7} is associated with the turbulence at the unstable interface. 

\begin{figure}
\includegraphics[scale=0.85,clip]{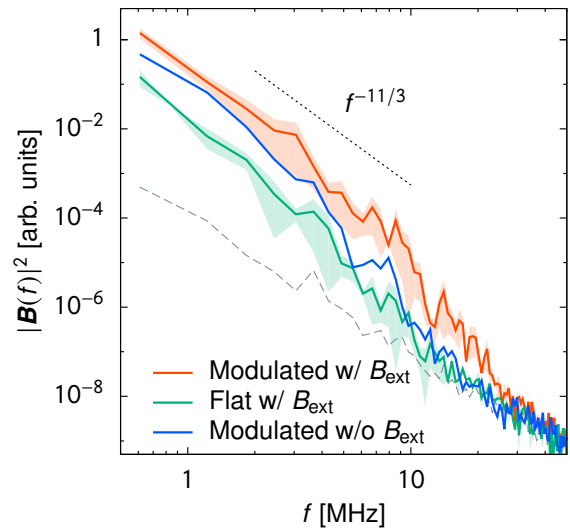}%
\caption{
Frequency spectra of the magnetic field energy calculated by the Fourier transform of the B-dot data for three types: (red) the modulated-target shots with the seed magnetic field, (green) the flat-target shots with the magnetic field, and (blue) the modulated-target shots without the magnet.
Each spectrum is the average of two different shots with the same experimental conditions. 
{\color{black}
The shot-by-shot fluctuation is indicated with the light color for the modulated-target (light-red) and flat-target (light-green) cases with the magnet.  
{\color{black}
The light-color thickness stands for the deviation from the average for each shot.}}
The selected period for the Fourier analysis is from 1.93 to 3.57 $\mu$s (see Fig.~\ref{fig7}).
The reference slope proportional to $f^{-11/3}$ is for the Kolmogorov turbulence. 
The dashed gray curve indicates the noise level calculated from the reference data taken before the shot.
  \label{fig8}}
\end{figure}

Figure~\ref{fig8} shows the Fourier spectra of the time variation of the B-dot signals for the three types shown in Fig.~\ref{fig7}.
The frequency dependence of the mode amplitude is depicted where the correction of the sensitivity is applied based on the calibration using the controlled time variation of magnetic fields. 
Each spectrum is the average of two-shot data with the same experimental conditions.
The shot numbers used in the B-dot analysis are listed in Table~\ref{tab1} and \ref{tab2} (see the last column of these tables).
{\color{black}
The shot-by-shot fluctuations are shown by thick lines with the light color in Fig.~\ref{fig8} for the modulated-target and flat-target cases with the magnet. 
The width indicates the deviation from the average. 
The relative deviation to the average is around 0.7 in a range from 1 to 10 MHz for all three cases.}

For the Fourier analysis, we concentrate on the contribution of the interface fluctuations.
Then, the B-dot data for the spectra are extracted between $t = 1.93$ and 3.57 $\mu$s as highlighted in Fig.~\ref{fig7}.
We measured the reference data before every shot. 
The noise level plotted in Fig.~\ref{fig8} is the average of the Fourier amplitude of the corresponding reference data. 
The B-dot signals of the shots are significantly higher than the noise level at a frequency of less than about 30 MHz, where a few tens of MHz is the diagnostic limitation of the B-dot coil.

The frequency spectrum for the modulated-target shot with the magnet has the highest amplitude.
This is the case of the RMI growth with a seed field.
The power index of the amplified magnetic field is close to the Kolmogorov value of $-11/3$ \cite{biskamp03}.
The enhancement of the magnetic field compared to the flat-target shot is seen at a frequency of less than 30 MHz.
The amplification factor is one order of magnitude larger in terms of the magnetic energy.
The Fourier spectrum for the self-generated field case appears in between the other two cases.
The ordering of the mode amplitude among these three types is reproducible and very general. 
The qualitative behavior of the magnetic fields measured by the B-dot probe is consistent with the radiation MHD simulations including the Biermann battery effect using the FLASH code \cite{fryxell00,calder02} (see Appendix~\ref{secb}).

If a constant speed of the plasma flow is assumed, the frequency information is replaced by the spatial size of the magnetic field fluctuations.
In other words, the horizontal axis of Fig.~\ref{fig8} can be regarded as the wavenumber of the fluctuations.
The frequency of 30 MHz corresponds to 300 $\mu$m when the plasma velocity is 10 km/s.
Consequently, the turbulent structure of the RMI would be larger than a few hundreds of $\mu$m, which is of the order of the initial modulation wavelength.
The B-dot signals at the later phase continue over 1 $\mu$s, so that the corresponding plasma size is more than 1 cm.
The entire region of the CH plasmas would be in a turbulent state when it reaches the location of the probe.
On the other hand, the contribution of the self-generated field is evident at $f \lesssim 3$--10 MHz, so that the spatial size may be larger than 1--3 mm for the Biermann effect.

\section{Discussion \label{sec4}}

\subsection{Interfacial instabilities}

The growth velocities measured in this experiment are consistent with the linear growth velocity of the RMI.
However, if the deceleration of the interface is not negligible, the exponential growth of the RTI influences the modulation amplitude.
Here, we estimate the growth rate of the RTI based on the experimental data.

The averaged interface velocity evaluated from the front position of the interface at $t = 40$ ns is $\langle v_i \rangle = 44.1$ km/s in the highest intensity shot. 
On the other hand, the temporal average from $t = 30$ to 40 ns is fitted as 38.7 km/s in the same shot.
Then the deceleration of the interface is roughly estimated by $g \sim -3.6 \times 10^{8}$ km/s$^2$, and the corresponding growth rate of the RTI is $({\rm At} g k)^{1/2} \sim 1.2 \times 10^8$ s$^{-1}$, where the Atwood number ${\rm At} \sim - 1 $ is assumed.
Therefore, the contribution of the RTI could appear after a few tens of nanoseconds, which is comparable to the observed timescale in our experiment. 
The deceleration is reduced slightly in the lower intensity cases. 
This picture explains the deviation from the theoretical growth velocity of the RMI at the higher intensity shown in Fig.~\ref{fig4}.

\subsection{Hydrodynamic Similarities}

Hydrodynamic and MHD phenomena are scalable from a small size in the laboratory to astronomical scale in the Universe \cite{remington99,ryutov99,ryutov00} (see Table~\ref{tab3}).
Thus, our experiment mimics the interstellar turbulence and has successfully demonstrated the amplification of the magnetic field by the interfacial instabilities in astrophysical events.
If the magnetic pressure is much smaller than the thermal pressure, the magnetic field has no impact on the RMI growth.
The evolution of the hydrodynamic RMI is characterized by three non-dimensional ratios, which are the shock velocity to the sound speed, the density ratio at the interface, and the corrugation amplitude to the wavelength, whatever value the denominator and numerator take. 

Let us consider the scaling relation between our RMI experiment and the supernova remnant (SNR).
Assuming the characteristic shock velocity, shock radius, and density of SNRs as $U \sim 10^4$ km/s, $R \sim 3 \; {\rm pc} \sim 10^{17}$ m, and $n \sim 1$ cm$^{-3}$, the corresponding physical quantities are evaluated by the hydrodynamic scaling from the experimental values.
The equivalent time and temperature are resulted in $t \sim 300 \; {\rm yr} \sim 10^{10}$ s and $T \sim 30$ keV, which are in a reasonable range for SNRs.
The fluctuation length and velocity in SNRs are scaled to $\lambda \sim 0.05$ pc and $\delta v \sim 300$ km/s.
Then, our experiment simulates the turbulent structure of the size of molecular cloud cores \cite{bergin07}.
The fluctuations of sub-pc size might evolve into the birthplace of stars after the radiative cooling \cite{hennebelle19}.
Thus, what we observed in the experiment is said to be a part of stellar recycling processes in the Universe.

The scaling cannot hold if the effects of collisional processes have to be considered.
In our case, the collision effect appears predominantly in the ohmic dissipation for the magnetic-field evolution.
If the non-dissipative assumption is valid, the magnetic field is amplified by the turbulent motions of the RMI \cite{sano12}. 
However, the dissipation could weaken the amplified magnetic field significantly.

\begin{table*}
\caption{
Scaling between the plasmas in laser laboratories and SNRs.
Here, 1 pc = $3.09 \times 10^{16}$ m and 1 yr = $3.15 \times 10^7$ s.
The fluctuation velocity and length are represented by the growth velocity and wavelength of the RMI, respectively.
For the kinematic viscosity for SNRs, the magnetized viscosity $\nu \; [{\rm m}^2/{\rm s}] = r_L v_{\rm th} = \bar{T}_i / B$ is adopted \cite{ryutov99}, where $r_L$ and $v_{\rm th}$ are the Larmor radius and thermal velocity of ions.
\label{tab3}}
\begin{tabular}{llrlrl}
\hline \hline
& Definition & \multicolumn{2}{c}{Laser-shocked Plasma}
& \multicolumn{2}{c}{SNR} \\
\hline 
Material & & CH & & H & \\
Mass Density & $\rho$ & 1 & g/cm$^3$
& $2 \times 10^{-24}$ & g/cm$^3$ \\
Electron Number Density & $n$ & $3 \times 10^{23}$ & cm$^{-3}$
& 1 & cm$^{-3}$ \\
Temperature & $\bar{T}$ & 10 & eV
& 30 & keV \\
Thermal Pressure & $P$ & $5 \times 10^{11}$ & Pa
& $5 \times 10^{-9}$ & Pa \\
Time & $t$ & 100 & ns
& 300 & yr \\
Shock Velocity & $U$ & 100 & km/s
& $10^4$ & km/s \\
Shock Radius & $R = Ut$ & 1 & cm
& 3 & pc \\
Plasma Velocity & $v$ & 30 & km/s
& $3 \times 10^3$ & km/s \\
Plasma Length & $L = vt$ & 3 & mm
& 1 & pc \\
Fluctuation Velocity & $\delta v$ & 3 & km/s
& 300 & km/s \\
Fluctuation Length & $\lambda$ & 150 & $\mu$m
& 0.05 & pc \\
Kinematic Viscosity & $\nu$ & $3 \times 10^{-7}$ & m$^2$/s
& $3 \times 10^{13}$ & m$^2$/s \\
Reynolds Number & Re = $\delta v \lambda / \nu$ & $2 \times 10^6$ &
& $2 \times 10^7$ &\\
Magnetic Diffusivity & $\eta$ & 30 & m$^2$/s
& $2 \times 10^{-4}$ & m$^2$/s \\
Magnetic Reynolds Number & Rm = $\delta v \lambda / \eta$ & 0.02 &
& $3 \times 10^{24}$ \\
Magnetic Prandtl Number & Pm = $\nu / \eta$ & $10^{-8}$ &
& $2 \times 10^{17}$ & \\
Magnetic Field & $B$ & 0.1 & T
& 1 & nT \\
Magnetic Pressure & $P_{\rm mag}$ & $4 \times 10^3$ & Pa
& $4 \times 10^{-13}$ & Pa \\
Alfv{\'e}n Speed & $v_A$ & 3 & m/s
& 20 & km/s \\
Plasma Beta & $\beta = P / P_{\rm mag}$ & $10^8$ &
& $10^4$ &\\
Alfv{\'e}n Number & Al = $\delta v / v_A$ & $10^3$ &
& 15 & \\
    \hline \hline
  \end{tabular}
\end{table*}

\subsection{Magnetic-field amplification}

The importance of the magnetic dissipation is usually indicated by the magnetic Reynolds number, which is defined by ${\rm Rm} = {\cal{V L}} / \eta$ using the characteristic velocity $\cal{V}$ and length $\cal{L}$. 
Here, $\eta = (\mu_0 \sigma)^{-1}$ is the magnetic diffusivity, $\sigma = e^2 n_e / (m_e \nu_{ei})$ is the electrical conductivity, $e$ is the elementary charge, $n_e$ and $m_e$ are the number density and mass of electrons, and $\nu_{ei}$ is the electron-ion collision frequency.
Using the Spitzer formula \cite{chen84}, the collision frequency is given by
\begin{equation}
  \nu_{ei} =
  \frac{\ln \Lambda}{3 (2 \pi)^{3/2}}
  \frac{Z e^4}{\varepsilon_0^2 m_e^{1/2}}
  \frac{n_{e}}{(k_B T_e )^{3/2}} \;,
\end{equation}  
where $\ln \Lambda \; (\sim 10)$ is the Coulomb logarithm, $Z \; (\sim 1)$ is the ion charge, $\varepsilon_0$ is the vacuum permittivity, and $T_e$ is the electron temperature. 
Considering the case of ${\cal{V}} \sim \delta {v}$ and ${\cal{L}} \sim \lambda$, it takes 
\begin{eqnarray}
  {\rm Rm} & \approx &
  \frac{e^2 n_e}{\varepsilon_0 m_e c^2 \nu_{ei}} {\cal{V L }} 
  \nonumber \\
  & \sim & 0.02
  \left( \frac{\bar{T}_e}{10 \; {\rm eV}} \right)^{3/2}
  \left( \frac{\delta v}{3 \; {\rm km/s}} \right)
  \left( \frac{\lambda}{150 \; \mu{\rm m}} \right)
   \;,
\label{eq:rem}
\end{eqnarray}
for our experimental conditions.
The typical temperature of the laser-shocked CH is adopted for $T_e$ \cite{barrios10}, and $\bar{T}$ is the temperature in eV.

This estimation tells us that the magnetic Reynolds number in the laser plasmas could be much smaller than that for astrophysical plasmas (see Table~\ref{tab3}).
The dissipation timescale is $\lambda^2 / \eta \sim 1$ ns for the parameters in Eq.~(\ref{eq:rem}), so that the saturation level of the turbulent magnetic field is determined by the balance between the amplification and ohmic dissipation. 
The low Rm might be the reason why the amplification factor is reasonably smaller than the result of ideal MHD simulations \cite{sano12}.
Nonlinear simulations, including ohmic dissipation, are inevitable for more quantitative discussions on the magnetic field. 

On the other hand, the fluid viscosity is negligible in our experiment.
The ion-ion collision frequency is written as $\nu_{ii} = (m_e / m_i)^{1/2} (T_e / T_i)^{3/2} (Z^2 / \sqrt{2}) \nu_{ei}$, where $m_i$ and $T_i$ is the mass and temperature of ions.
The Reynolds number is defined as ${\rm Re} = {\cal{V L}} / {\nu}$ by using the kinematic viscosity $\nu = k_B T_i / (m_i \nu_{ii})$, which takes
\begin{eqnarray}
  {\rm Re} \sim 2 \times 10^6
  \left( \frac{\rho}{1 \; {\rm g/cm}^3} \right)^{-1}
  \left( \frac{\bar{T}_i}{10 \; {\rm eV}} \right)^{5/2} \nonumber \\
  \times \left( \frac{\delta v}{3 \; {\rm km/s}} \right)
  \left( \frac{\lambda}{150 \; \mu{\rm m}} \right)
  \;.
\end{eqnarray}
Here we use a relation for the ion density $n_i = \rho / (A m_p)$ where $A \; (\sim 6.5)$ is the mass number and $m_p$ is the proton mass.
The viscous timescale is much longer than the dissipation timescale, because the magnetic Prandtl number is quite small, ${\rm Pm} = \nu/\eta \sim 10^{-8}$.

\subsection{Self-generated magnetic fields}

The self-generated magnetic field will affect the saturation level of the field in the RMI turbulence.
It is found that the self-generated field has a detectable contribution in the B-dot signals.
The Biermann battery term in the induction equation is given by
\begin{equation}
  \frac{\partial \bm{B}}{\partial t} = \frac1{e n_e^2} \left(
  \nabla P_e  \times \nabla n_e \right) \;,
\end{equation}
where $P_e = n_e k_B T_e$ is the electron pressure.
The order of magnitude estimate of the self-generated field is written as
\begin{eqnarray}
  B_{\rm self} \; [{\rm T}] \; &\approx& \frac{P_e}{e n_e {\cal{V L}}} \nonumber \\
  &\sim& 2
 \left( \frac{\bar{T}_e}{10 \; {\rm eV}} \right)
  \left( \frac{v_i}{30 \; {\rm km/s}} \right)^{-1}
  \left( \frac{\lambda}{150 \; \mu{\rm m}} \right)^{-1}
   \;,
\end{eqnarray}
using the typical values of ${\cal{V}} \sim v_i$ and ${\cal{L}} \sim \lambda$ for the laser experiment. 
Note that $B_{\rm self}$ at SNRs is negligibly small compared to the ambient magnetic field, so that this is a unique feature of the laser RMI experiment.

{\color{black}
The kinematic viscosity is tiny in our situation (see Table~\ref{tab3}). 
Then, the velocity fluctuations initiated by the RMI could remain for much longer than several tenths of nanoseconds. 
As long as the turbulent motions exist, amplification and self-generation of the magnetic field can still happen.}
If the Biermann effect is the dominant mechanism of the field enhancement, the balance with the ohmic dissipation brings the saturation amplitude of the magnetic field, that is,
\begin{equation}
  B_{\rm sat} \; [{\rm T}] \approx \frac{P_e}{e n_e \eta} \sim 0.4
  \left( \frac{\bar{T}_e}{10 \; {\rm eV}} \right)^{5/2} 
 \;.
\end{equation}
The amplitude is determined only by the temperature for this case.
The saturated field strength is independent of the size and velocity of the turbulence, although they affect the timescale of saturation.

In our experiment, the Alfv{\'e}n number,
\begin{equation}
 {\rm Al} \sim 10^3
  \left( \frac{\rho}{1 \; {\rm g/cm}^3} \right)^{1/2}
  \left( \frac{B}{0.1 \; {\rm T}} \right)^{-1}
  \left( \frac{\delta v}{3 \; {\rm km/s}} \right) \;,
\end{equation}
is always large enough to guarantee the passive evolution of magnetic fields by turbulent motions.
The Alfv{\'e}n number is also greater than unity for the SNR parameters in Table~\ref{tab3}.
The measurement of the field strength must be an essential next step.
Furthermore, {\it in situ} measurements of the density and velocity fluctuations in the RMI turbulence are worth challenging for the feedback in understanding the interstellar turbulence and star formation scenarios.

\subsection{Laser astrophysics experiments}

An exciting extension of this work is to confirm the suppression of the RMI by a strong magnetic field experimentally. 
The suppression and amplification processes can be understood continuously in terms of the size of the Alfv{\'e}n number ${\rm Al}$ \cite{sano13,sano21}. 
When the Alfv{\'e}n number is less than unity, the interface oscillates stably after the shock passage. 
The required strength for the suppression is larger than $B_{\rm crit} \sim 100$ T for typical laser-plasma conditions as given by Eq.~(\ref{eq:bcrit}). 

At present, strong magnetic fields of kilo-Tesla order are available in the laser experiments by several methods \cite{yoneda12,fujioka13,korneev15,goyon17}.
By introducing capacitor coil targets to generate a quasi-static magnetic field over 100 T \cite{fujioka13}, we could examine the suppression regime of the RMI in the same experimental setup using high-power laser facilities.
The lower density target reduces the critical field strength so that the RMI could be mitigated by a more easily manageable condition for the external magnetic field.
In this sense, it would be interesting to use a modulated foam target surrounded by the gas for this purpose.

The dependence of the RMI growth on the direction of the initial magnetic field is another interesting topic for future laser experiments.
In our setup, the initial field $B_z$ is amplified by the RMI motions.
For the suppression study, the field direction distinguishes the final state of the RMI.
The $x$ and $z$ components work as the suppression force on the RMI.
Thus, the strong $B_x$ and $B_z$ could reduce the growth of the RMI.
However, if the initial field has only $y$ component in our setup, which is perpendicular to the RMI motions, the magnetic field cannot stabilize the RMI at all.
This kind of multi-dimensional effect may have a significant meaning for the application to the implosion process in laser-driven ICF plasmas \cite{hohenberger12,wang15,fujioka16,perkins17}.

\section{Conclusions \label{sec5}}

We have investigated the amplification of a seed magnetic field by the growth of the RMI associated with a laser-driven shock wave.
Our findings are summarized as follows: 

\begin{enumerate}
\item
The unstable growth of the surface corrugation is captured by the optical shadowgraph in our laser-induced shock experiment.
The growth velocity observed in the experiment is consistent with the linear growth velocity predicted by the analytical theory of the RMI.
However, when the laser intensity is higher, the contribution of the RTI enhances the fluctuation amplitude in addition to the RMI.
\item
The induction coil probe successfully measures the evidence of the magnetic-field amplification by the RMI.
It is found that the random field in the RMI turbulence has the spatial structure of the order of the initial RMI wavelength.
The saturation level of the magnetic field would be determined by the balance between the turbulent amplification and ohmic dissipation in our experiment.
\item 
When the RMI takes place, the signals of magnetic fields are always detected with or without a seed field.
It confirms that self-generated fields through the Biermann battery process are non-negligible in the RMI turbulence for the laser-plasma case.  
\end{enumerate}

This work is primarily motivated to understand the evolution of interstellar turbulence and magnetic fields.
Magnetic-field generation and amplification by the interfacial instabilities are demonstrated distinctly in our laser experiment.
The coupling with ambient magnetic fields in interstellar plasmas is stronger than that in laboratory laser plasmas, and thus the field amplification by turbulent motions occurs undoubtedly in many astrophysical phenomena.
Therefore, the RMI must have a significant contribution to the emergence of strong magnetic fields associated with supernova shocks.
This fundamental research will be applicable to various subjects other than astrophysics.
For instance, the MHD behavior of the RMI is crucially important to the optimization of the implosion process for laser-driven ICF.

\begin{acknowledgments}
This work was performed under the joint research project of the Institute of Laser Engineering, Osaka University. 
We are deeply grateful to the GEKKO technical crew for their exceptional support during these experiments.
We thank F. Cobos-Campos and J. G. Wouchuk for the linear analysis of the RMI. 
We also thank M. Hoshino, K. Katagiri, Y. Kuramitsu, S. Matsukiyo, N. Ozaki, and R. Yamazaki for useful discussion, and J. L. Gabayno for her careful reading of the manuscript.
The software used in this work was in part developed by the DOE NNSA-ASC OASCR Flash Center at the University of Chicago. 
This research was partially supported by JSPS KAKENHI Grant No. JP26287147, No. JP15H02154, No. JP16H02245, and No. JP19KK0072, JSPS Core-to-Core Program, B. Asia-Africa Science Platforms No. JPJSCCB20190003, and MEXT Quantum Leap Flagship Program Grant No. JPMXS0118067246.
This work was also supported by the Agence Nationale de la Recherche (ANR) in the framework of the ANR project TURBOHEDP (ANR-15-CE30-0011). 
\end{acknowledgments}

\appendix

\section{LINEAR GROWTH VELOCITY OF RMI \label{seca}}

The RMI is triggered by the deposition of the circulation when an incident shock passes through a corrugated density interface. 
Because of the corrugation of the transmitted and reflected wavefronts, the tangential velocities are generated by the refraction motions.
Then, the difference in the pressure fluctuations appears across the interface, which could be the driving force of the instability. 

The detailed linear theory of the RMI has been done in the form of series expansions in terms of the Bessel functions \cite{wouchuk97}.
Consider an interaction of a corrugated interface between two fluids (``a'' and ``b'') and a planner shock traveling in the fluid ``b''.
The asymptotic growth velocity is calculated with the following expression: 
\begin{equation}
  v_{\rm wn} =
  \frac{- \rho_{a}^{\ast} \delta v_{a}^{\ast}
        + \rho_{b}^{\ast} \delta v_{b}^{\ast}}
       {\rho_{a}^{\ast} + \rho_{b}^{\ast}} +
  \frac{\rho_{a}^{\ast} F_a - \rho_{b}^{\ast} F_b }
       {\rho_{a}^{\ast} + \rho_{b}^{\ast}} \;, 
\label{eq:dvi}
\end{equation}
where $\rho_{a}^{\ast}$ ($\rho_{b}^{\ast}$) and  $\delta v_{a}^{\ast}$ ($\delta v_{b}^{\ast}$) are the density and tangential velocity at the interface of the fluid ``a'' (``b'') just after the shock passage. 
The quantity $F_a$ ($F_b$) represents the sonic interaction between the contact surface and the transmitted (reflected) wavefront, which are measured by the amount of vorticity left behind the wavefront in the bulk of each fluid.

The WN formula given by Eq.~(\ref{eq:dvi}) is exact within the limits of linear theory and inviscid flow. 
It is valid for any initial configuration, and every element is analytically calculated from the pre-shocked parameters \cite{wouchuk01b,coboscampos17}.
The growth velocity $v_{\rm wn}$ is determined by a given set of the parameters, which are the Mach number of the incident shock $M$, the pre-shocked density jump $\rho_{a0} / \rho_{b0}$ and the sinusoidal modulation amplitude relative to the wavelength $\psi_0 / \lambda$, and the isentropic index of the fluid $\gamma$.   
The first term of the right-hand side of Eq.~(\ref{eq:dvi}) is due to the instantaneous deposition of the vorticity at the interface just after the shock interaction, which has the dominant contribution in the limit of weak incident shocks.
On the other hand, the second term becomes non-negligible for stronger shocks or highly compressible fluids, and usually has the opposite sign to the first term.

\begin{figure}
  \includegraphics[scale=0.85,clip]{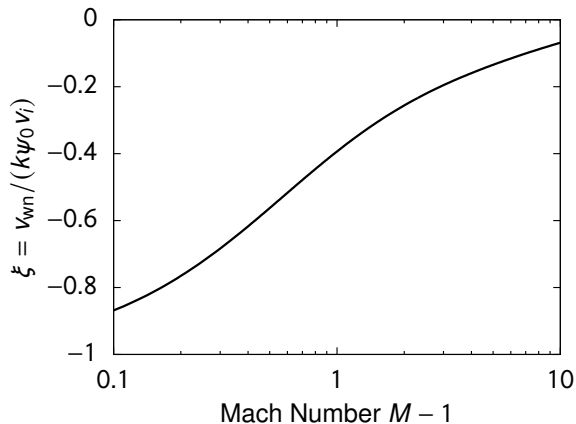}%
  \caption{
Mach-number dependence of the linear growth velocity of the RMI.
The relation between the growth velocity of the WN model $v_{\rm wn}$ and the interface velocity $v_i$ shown as a function of the incident Mach number $M$.
The vertical axis is the ratio defined by a non-dimensional factor $\xi = v_{\rm wn} / (k \psi_0 v_i)$.
The experimental parameters are adopted here for the evaluation of $v_{\rm wn}$, which are the density jump $\rho_{a0} / \rho_{b0} = 10^{-5}$ and the corrugation amplitude $\psi_0 / \lambda = 0.05$.
The isentropic index is assumed to be $\gamma = 5/3$.
\label{fig9}}
\end{figure}

Here we define a non-dimensional factor
\begin{equation}
  \xi = \frac{v_{\rm wn}}{k \psi_0 v_i} \;,
\end{equation}
where $k = 2 \pi / \lambda$ is the wavenumber of a mode.
Figure~\ref{fig9} shows the dependence of $\xi$ on the incident Mach number $M$ under our experimental conditions, where the density jump is $\rho_{{\rm N}_2}/\rho_{\rm CH} \approx 10^{-5}$ and the corrugation amplitude is $\psi_0 / \lambda \approx 0.05$.
If the equation of state for the ideal gas with $\gamma = 5/3$ is assumed, the ratio of the growth velocity of the WN model to the interface velocity, $v_{\rm wn} / v_i$, is determined only by the Mach number $M$. 
The negative velocity stands for the phase reversal that is a typical feature of the RMI for the rarefaction-reflected cases. 

As can be seen from Fig.~\ref{fig9}, the factor $\xi$ ranges from $-0.4$ to $-0.1$ for the strong shock limit of $M \gtrsim 2$.
It is difficult to define the incident Mach number in our experiment based only on the observable optical information.
Here we adopt $\xi \sim -0.3$ (around $M \sim 3$) for the estimation of the growth velocity.
In the end, the theoretical growth velocity of the WN model is depicted in Fig.~\ref{fig4} with the help of the observed interfacial velocity given by Eq.~(\ref{eq:vint}). 

\section{NUMERICAL SIMULATIONS ON 
  AMPLIFICATION AND SELF-GENERATION OF MAGNETIC FIELDS \label{secb}}

\begin{figure*}
  \includegraphics[scale=0.85,clip]{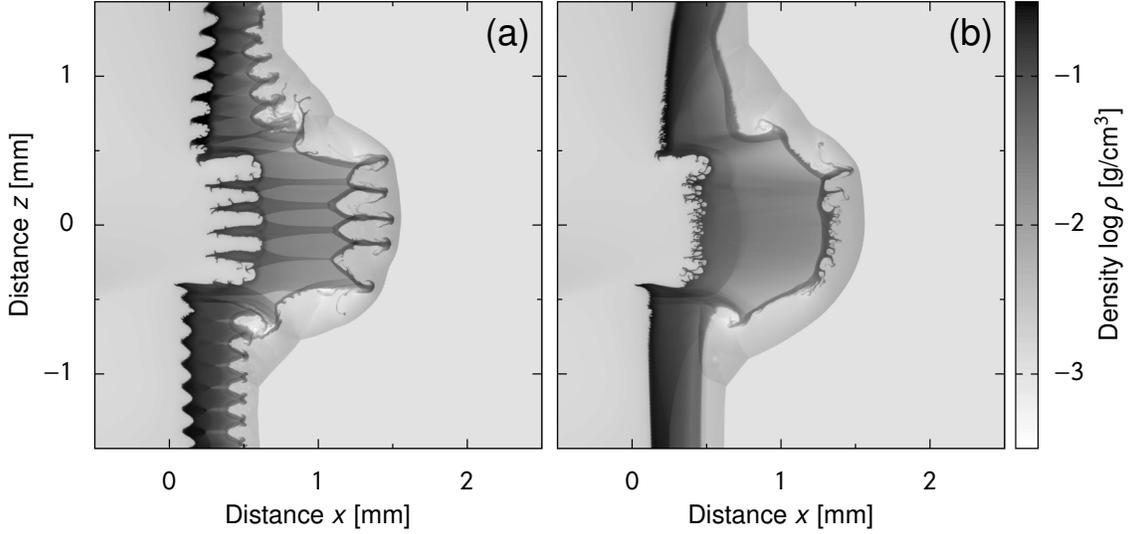}%
\caption{
Snapshots of the density distribution at 50 ns after the laser irradiation calculated by the FLASH code.
The initial conditions of these radiation MHD simulations are almost the same as the experimental parameters for the cases of (a) a modulated target and (b) a flat target.
Finger-like structures of the interface indicate the growth of the RMI with the wavelength of the initial modulation.
Although there are tiny fluctuations of the interface in the flat-target case, the contact discontinuity is relatively smooth, the same as seen in the experiment.
\label{figs4}}
\end{figure*}

Experimental evidence of the magnetic-field amplification comes from the B-dot probe.
The unstable motions driven by the RMI amplify the ambient magnetic field by stretching and compressing field lines, which has been confirmed by the ideal MHD simulations assuming the single-mode analysis \cite{sano12}.
It is predicted that the amplification factor can reach two orders of magnitude or more.
Although the geometrical effects of spherical expansion may reduce the field strength in the actual experiment, the growth of the RMI could bring detectable differences between the cases with the modulated target and flat target.
However, self-generated magnetic fields should be considered in the numerical study relevant to the laser experiment. 
Thus, to estimate the evolution of magnetic fields in our experiment, we performed radiation MHD simulations using FLASH code \cite{fryxell00,calder02} including the Biermann battery term. 

As the initial conditions for the numerical simulations, we adopt a similar configuration to our experiment.
A modulated CH foil with the density 1 g/cm$^3$ is put in the atmospheric helium gas.
The Cartesian coordinate in two-dimensions ($x$, $z$) is used, where the $x$ and $z$ directions are perpendicular and parallel to the target surface. 
The target thickness is 50 $\mu$m, and the location of the front surface is at $x = 0$.
We prepare two types of targets, which are a modulated target and a flat target.
For the modulation at the rear surface, the wavelength of the sinusoidal pattern is 150 $\mu$m with an amplitude of 8.8 $\mu$m.  
The gas density is chosen to be $10^{-3}$ g/cm$^3$ from the constraint of numerical computation, which is slightly denser than in the experiment.
A uniform magnetic field is applied in the direction of 45 degrees to the target surface, $B_x = - B_z > 0$.
The initial field strength is 0.1 T.
In terms of the laser conditions, the pulse shape is a square wave of 2.5 ns.
The incident angle of the laser injection is 45 degrees to the target surface, and it is normal to the direction of the seed magnetic field. 
The spot size is 600 $\mu$m, and the laser intensity corresponds to $1.1 \times 10^{13}$ W/cm$^2$ at the target surface.
The center of the laser focal spot is set to be at the origin, $(x, z) = (0, 0)$. 

The range of the computational domain is sufficiently larger than the spot size, that is, $-800$ $\mu${\rm m}$ \le x \le 3200$ $\mu{\rm m}$ and $|z| \le 1500$ $\mu{\rm m}$.
The outflow boundary conditions are assumed at all four boundaries. 
An adaptive mesh refinement technique is adopted to capture narrow structures of the vortex at the interface.
The grid size is determined according to the magnitude of the density and temperature gradients, and the minimum grid size in our simulations is 0.98 $\mu$m. 

\begin{figure*}
  \includegraphics[scale=0.85,clip]{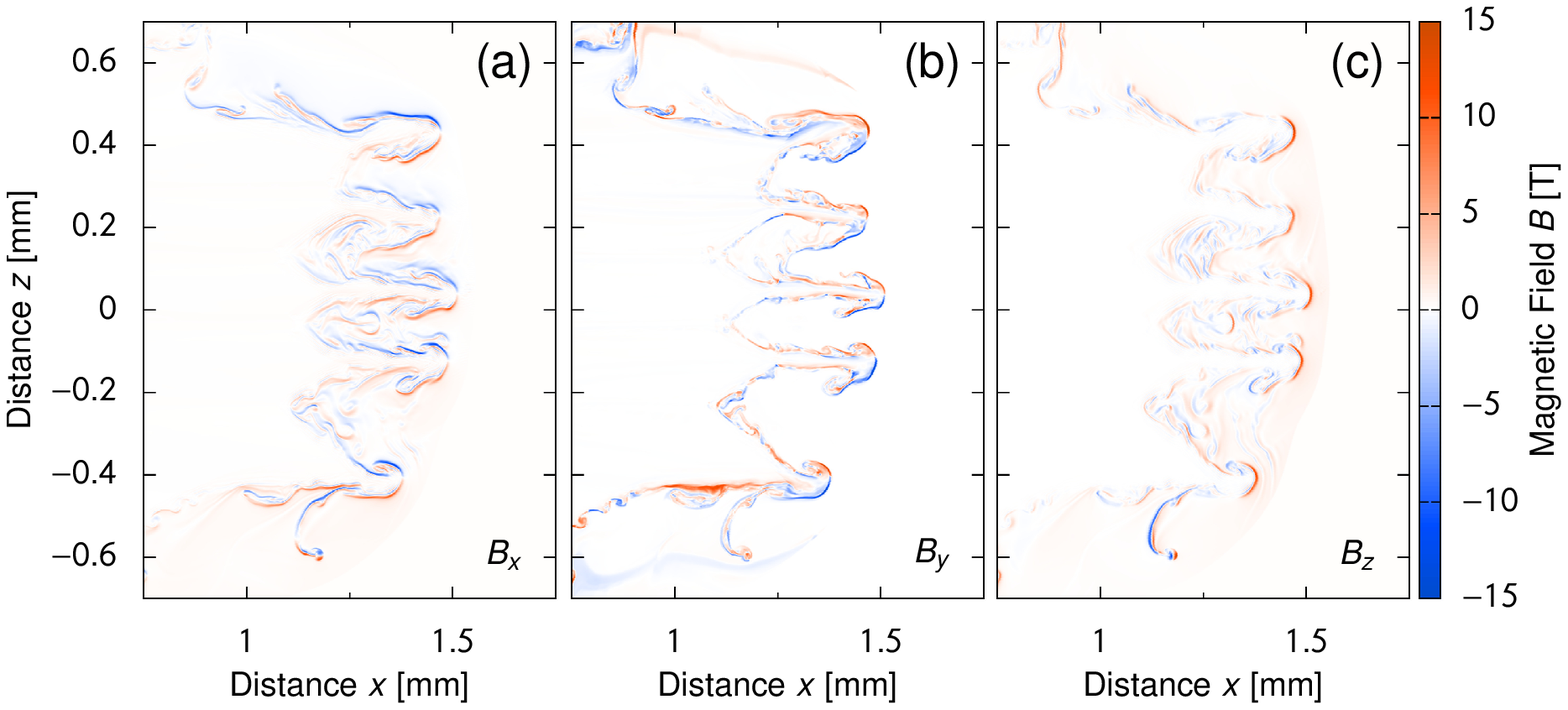}%
\caption{
Spatial distributions of the magnetic field at the nonlinear regime of the RMI in the modulated-target case.
The color denotes the strength of the magnetic field in the unit of Tesla for (a) $B_x$, (b) $B_y$, and (c) $B_z$.
The snapshots are taken at 50 ns after the laser hits the target.
The $x$ and $z$ components are amplified by the RMI growth from a weak seed field.
On the other hand, the $y$ component is generated only through the Biermann battery effect. 
\label{figs5}}
\end{figure*}

\begin{figure*}
  \includegraphics[scale=0.85,clip]{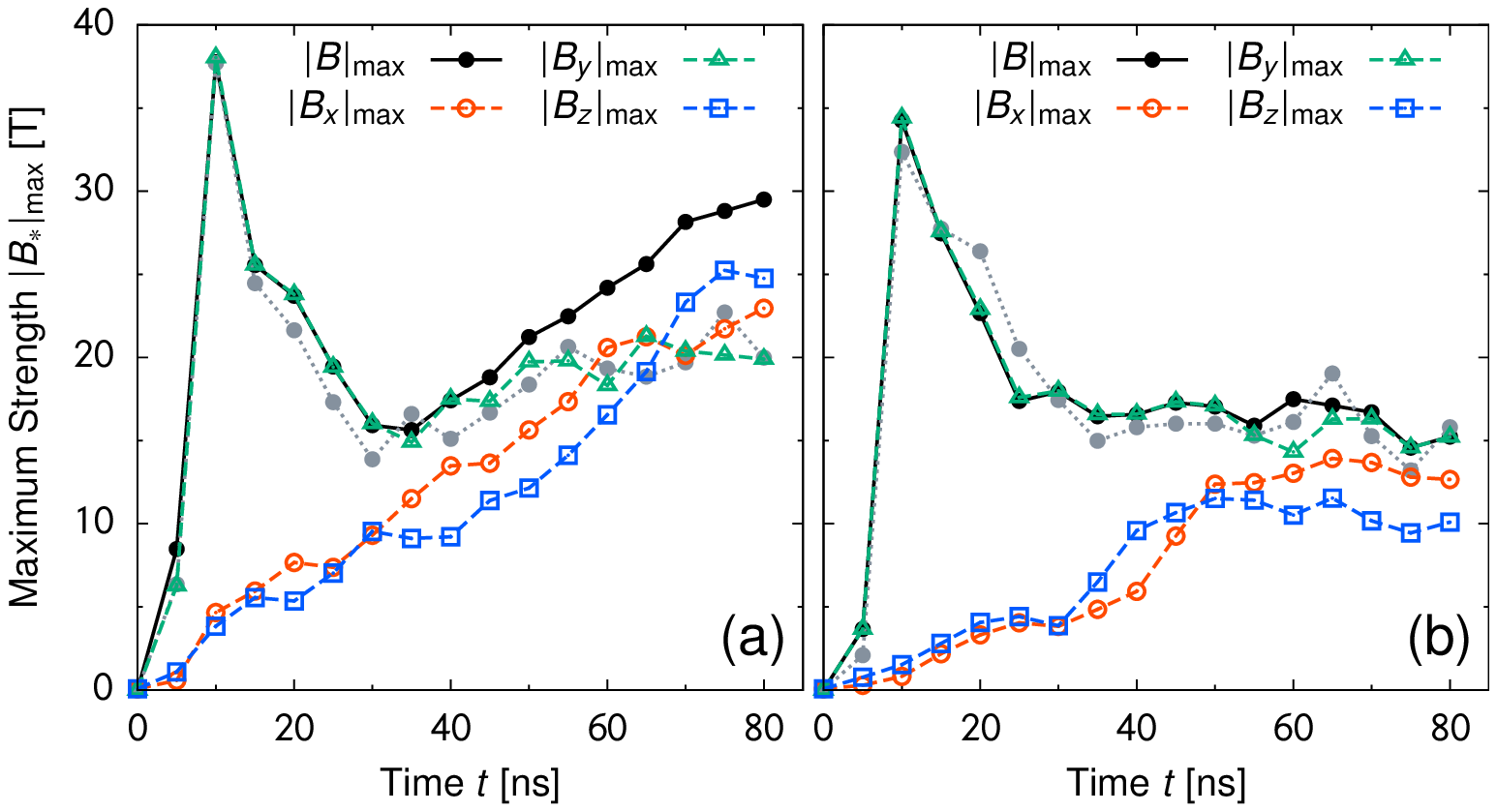}%
\caption{
Time evolutions of the maximum strength of the magnetic field for each component in (a) the modulated-target case and (b) the flat-target case.
The laser irradiation is from $t = 0$ to 2.5 ns.
The amplified magnetic fields $|B_x|_{\max}$ and $|B_z|_{\max}$ are depicted by the red circles and blue squares, respectively.
The green triangles indicate the self-generated magnetic field $|B_y|_{\max}$.
The time profiles of the self-generated magnetic field in the reference simulations without the initial magnetic field are also shown by the gray circles.
\label{figs6}}
\end{figure*}

Figure~\ref{figs4}(a) shows a snapshot of the density distribution at 50 ns after the laser irradiation. 
The growth of the RMI triggered by the shock passage is recognized as the interface fluctuations.
Several finger-like structures with about 300 $\mu$m long are formed, which exhibits obvious difference from the flat-target shot depicted by Fig.~\ref{figs4}(b).
The interface velocity and the growth velocity of the fluctuation amplitude in these simulations are consistent with the experiment quantitatively. 
However, there are some discrepancies in the detailed structure of the finger shape and the shock front position compared with the experimental images.

The strong magnetic fields are observed near the fluctuated interface.
The magnetic field distributions of each component are shown in Fig.~\ref{figs5}.
Because of the uniformity in the $y$ direction, the amplified magnetic fields always have $B_x$ and $B_z$ components, whereas the self-generated magnetic fields appear only in $B_y$. 
Thus, the self-generated component can be distinguished completely from the amplified component.

The initial weak fields are amplified along with the interface by the stretching motions associated with the RMI growth.
The shock compression also contributes to the magnetic-field enhancement of the $z$ component [see Fig.~\ref{figs5}(c)].
The maximum strength of the amplified magnetic field is around 10 T in this simulation, which is about 100 times larger than the initial field. 
These features in the amplified magnetic fields are consistent with the ideal MHD cases \cite{sano12}.

On the other hand, magnetic fields can be generated through the Biermann battery effect without any seed fields.
The large Biermann fields are caused by the large vorticity so that the strong $B_y$ also appears along with the interface [see Fig.~\ref{figs5}(b)].
The maximum strength of the self-generated magnetic field is comparable to the amplified magnetic field at the time of the snapshot.
In the experiment, we observed a mixture of the amplified and generated magnetic fields.

It is interesting to compare the maximum field strength between the modulated-target and the flat-target cases. 
The time evolutions of the maximum field strength for each component are plotted by Fig.~\ref{figs6}.
Here, the maximum value is searched within the range of the laser spot, $|z| \le 200$ $\mu$m.

The self-generated $B_y$ takes a peak value at the early phase of the evolution for both cases. 
The peak value of the self-generated magnetic field is about 35 T, which is much larger than the initial ambient field of 0.1 T.
However, the self-generated magnetic field decreases shortly within a few tens of nanoseconds.
In contrast, the ambient magnetic field is gradually amplified associated with the growth of the RMI. 
The maximum strength of the amplified components exceeds the self-generated $B_y$ sufficiently after the laser shot around $t \gtrsim 60$ ns.  
In the flat-target simulation, the time history of the self-generated magnetic field is similar to that for the modulated-target case, because this is an ablation-side phenomenon.
The saturated strength of $B_y$ determined in the rear-side plasmas is slightly weaker in the flat-target case.  
As can be seen from Fig.~\ref{figs4}(b), even in the case of the flat target, there is some disturbance growth at the interface, which may be originated from the nonuniformity of the laser absorption or numerical noise of the grid-size scale.  
The seed magnetic field is amplified by this small interfacial perturbation. 
However, the amplified magnetic field is much weaker than that in the RMI case, and it never reaches the strength of the self-generated magnetic field. 

Since the initial magnetic field is too weak to affect the dynamical evolution of the RMI, the time evolution of the self-generated magnetic field is almost unchanged by the presence of the initial magnetic field. 
Then, the $B_y$ profile alone can be regarded as the magnetic-field evolution for the cases without the magnet.
This interpretation is confirmed by the simulation results without the initial magnetic field shown in Figs.~\ref{figs6}(a) and \ref{figs6}(b) by the gray circles.
The experimental data indicate that the Fourier amplitude of the magnetic energy for the modulated target with the magnet is larger than that for the modulated target without the magnet.  
Thus, the characteristics of the magnetic fields in the numerical simulations are consistent with the experimental fact. 
In the experiment, the weakest magnetic field was measured in the flat target case, which is also reproduced correctly by the simulations.

The experiments and simulations show good agreement with respect to the relative strength of the magnetic field for three different types shown in Figs.~\ref{fig7} and \ref{fig8}.
Thus the MHD simulations support the positive correlation between the RMI growth and the magnetic-field amplification observed in our experiment.
Based on the simulations, it is implied that when a flat target is used without a magnet, the B-dot signal would not be so different from that in a flat-target shot with a magnet. 
In the experiment, the B-dot probe measures the magnetic field at a much later time than in the simulations.
During the long-term evolution, the magnetic field could be affected by magnetic dissipation and three-dimensional geometrical effects. 
Therefore, more extended radiation MHD simulations, together with quantitative measurements of the magnetic field in experiments, will be essential for future studies.


%


\end{document}